# Exploring Deep Learning Methods for Classification of SAR Images: *Towards NextGen Convolutions via Transformers*

Aakash Singh & Vivek Kumar Singh

*Abstract*— Images generated by high-resolution SAR have vast areas of application as they can work better in adverse light and weather conditions. One such area of application is in the military systems. This study is an attempt to explore the suitability of current state-of-the-art models introduced in the domain of computer vision for SAR target classification (MSTAR). Since the application of any solution produced for military systems would be strategic and real-time, accuracy is often not the only criterion to measure its performance. Other important parameters like prediction time and input resiliency are equally important. The paper deals with these issues in the context of SAR images. Experimental results show that deep learning models can be suitably applied in the domain of SAR image classification with the desired performance levels.

## I. INTRODUCTION

SAR stands for Synthetic Aperture Radar. It is a kind of active and coherent data collection system which is mostly airborne or spaceborne. It utilizes the long-range propagation properties of radar signals. It works by analyzing the strength of reflected signals after interacting with objects/ surfaces. It is employed in tasks that require imaging of broad areas at high resolutions. The application domain of this technique is very vivid containing areas like earth-resource monitoring, disaster management, and military systems. In such use cases, most of the time the light and the weather conditions remain unfavorable for optical imagery. However, the SAR technique attracts more noise than its optical counterparts. The noise here is called Speckle which comprises granular noise patterns that distort the quality of the image.

With the advancement in technology, high-resolution SARs are now coming into the picture. This has made the manual classification of the images a tedious and unremunerative task, thus highlighting the necessity of Automatic Target Recognition (ATR). The approaches used in ATR have been greatly influenced by machine learning techniques since their introduction. An ATR framework for SAR is composed of typically three phases: detection, discrimination, and classification [1]. The first phase evolves the detection of regions of interest (ROI) in an image. It is followed by the discrimination phase whose task is to filter out natural clutters. The last phase is the classification phase where we try to predict a label for the objects obtained from previous phases. This paper primarily deals with the classification phase. Some well-known state-of-the-art models are explored for their suitability in SAR image classification task. Thus, the paper presents an experimental study of application of deep learning models in the SAR image classification domain.

## II. RELATED WORK

Several previous studies have tried to look into the classification problem of SAR images. One such classification benchmark dataset (MSTAR) came in the late '90s. This led to researchers starting to work on this problem as early as 1998. One of the initial studies [2] used Bayesian pattern theory to deduce the importance of matching noise models to the noise statistics in classification. Later in 1999, another work [3] tried to explore the 3-class classification problem of MSTAR with polynomial Support Vector Machine (SVM) and managed to achieve an accuracy of 93.4%. A similar machine learning-based study [4] has applied SVM with the gaussian kernel on the same 3-class classification task and has compared the results with other traditional methods. Another research work [5] analyzed various feature extraction techniques namely Principal Component Analysis (PCA), Independent Component Analysis (ICA), and Hu moments, and clubbed it with popular classification techniques, and has touched classification accuracy of 98% on 8-class classification. The deep learning-based methods were seen to be employed in the domain in the year 2016. An AlexNet-based classifier network was created [6] and was modified with dropout layers. They called it "AFRLeNet" and have managed to score an accuracy of 99% on 7-class classification. Another research work [7] has applied deep convolutional neural networks (CNN) as the classifier. This work considered phase information along with amplitude information and claimed to reach an accuracy of 91% on 10-class classification. Few other studies also tried to further improve the accuracy of the classification. However, most of the studies ignored the parameter of responsiveness of the classification model, seen in terms of time required for prediction and the model resiliency. This work attempts to bridge this research gap by exploring the applicability of convolution and transformer-based models in SAR image classification.

## III. DATASET

This study utilizes the MSTAR (Motion and Stationary Target Acquisition and Recognition)[1] dataset. The dataset contains SAR radar images of targets (mainly military vehicles). It was collected and released in the public domain by Sandian National Laboratory[2]. The data was the outcome of a project jointly supported by Defense Advanced Research Projects Agency (DARPA) and the Air Force, USA. The collection of data was done in 2 phases i.e., in September 1995 (Collection #1) and in November 1996 (Collection #2). There are different versions of the datasets available. The version used in the study is the mixed variant of both the collection phases. It has SAR images of 8 different tank targets "Fig 1".

---

[1] https://www.sdms.afrl.af.mil/index.php?collection=mstar
[2] https://www.sandia.gov/

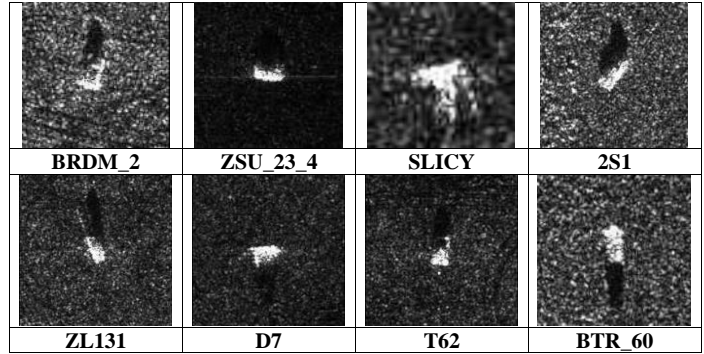

Fig. 1. Sample images from MSTAR dataset

The images generated for the same target at different depression angles may show wide variations. Hence the classification task is a challenging task. The distribution of data inside the classes is also visualized in "Fig. 2". It is done as the data distribution is known to largely affect the model training process and also in understanding the performance of the model in the evaluation stage [8].

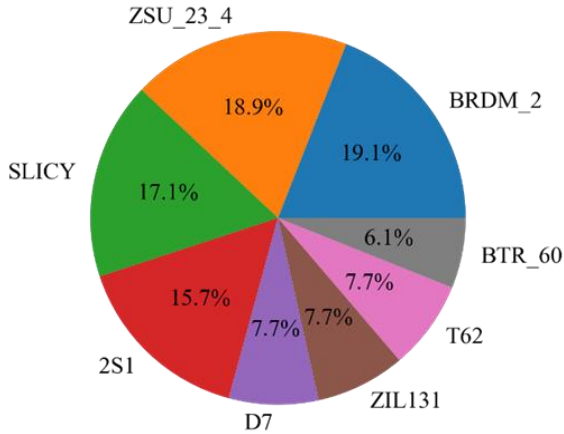

Fig. 2. Class distribution among dataset

IV. THE EXPLORED MODELS

The recent deep learning-based models used in the comparative analysis have been briefly described in this section. The pre-trained versions of these models were obtained from either the Keras applications library or from the TensorFlowHub. We have used base versions of these models to ensure an unbiased comparative study. The models were used in their vanilla form by fine-tuning their feature vectors only. The top layers of these models were neglected and instead, we added our dense prediction layer to suit the target classification needs.

*A. BiT*

It stands for Big Transfer, which is a term introduced by researchers of Google brains in their paper [9]. The purpose behind devising this technique was to tackle the problem of scarcity of labeled data. The standard method i.e., transfer learning has been known to struggle when applying patterns learned on a large dataset to smaller datasets. The model architecture of BiT employs standard ResNet with their depth and width increased. The paper discusses two important components that are crucial for transfer learning i.e., Upstream pre-training and Downstream fine-tuning. The first utilizes the fact that "larger datasets require larger architectures to realize benefits and vice-versa". The batch normalization of ResNet is replaced with GroupNorm and Weight Standardization (GNWS). For the second one, they have proposed their cost-effective fine-tuning protocol called "BiT-HyperRule". For our case, we have used BiT-M R50x1 version of the model pre-trained on the ImageNet-21k dataset available on TensorFlow Hub.

*B. ConvNext*

Since the introduction of transformers and their variants applicable to computer vision tasks, a lot of attention has been given by researchers to these models. This has even led to the negligence of mainstreamed convolutional neural network (CNN) based approaches. But, the recent paper [10] by researchers at Facebook AI Research(FAIR) has led to the revival of CNN. The researchers here have tried to improve the residual network-based architecture on the path of the hierarchical transformer. The ConvNext blocks "Fig. 3" are defined in terms of the transformer's self-attention block. ConvNext has been demonstrated to have rivaled the performance of most vision transformers. In addition, as they work on CNN backbone, they are not restricted by the size of the input image and can be trained feasibly on larger-size images. Out of several versions available we have used the base version of ConvNext downloaded from TensorFlow Hub.

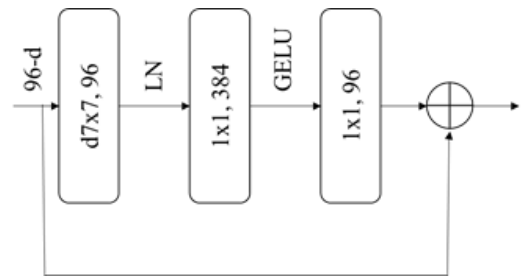

Fig. 3. ConvNext block

*C. DenseNet121*

DenseNets are among the most popular CNN-based architectures to be used in the computer vision domain. When first introduced in CVPR 2017 [11], it won the best paper award. It tends to harness the power of skip-connections, initially introduced for residual networks. Here, each layer is provided with an extra input made of feature maps of all the layers preceding it. To make feasible concatenation operation of previous inputs, the architecture is divided into Dense Blocks where the dimensions of the feature map do not change. There exist Transition layers between 2 Dense blocks. The DenseNet121 version of the architecture was chosen from Keras application library for this study.

*D. MobileNetV3*

This model was proposed keeping two key objectives in mind i.e., efficiency and light weightiness. The mobile applications were the target deployment of this model. It is also quoted as "TensorFlow's first mobile computer vision model" [12]. The proposed architecture uses depth-wise separable convolutions which reduces the number of parameters by many folds when compared with the regular convolution of the same depth. For our analysis, we have used the V3 [13] version of MobileNet which was introduced much later in 2019 with significant improvements over the original structure. The pre-trained model feature extractors were loaded from TensorFlow Hub.

*E. ViT*

The transformer was initially developed for NLP problems in 2017 in the paper titled "Attention is all you need" [14]. Soon with its introduction, it became the new state-of-the-art in the NLP domain. Later in 2020, the work of Dosovitskiy et al. 2020 [15] showed that the concept may also be applicable in the computer vision domain. This led to the birth of the Vision Transformer (ViT). The concept proved that dependency on CNNs can be removed completely by the introduction of a pure self-attention-based mechanism. ViT works by dividing the image into patches. These patches are embedded with the position vectors. This is then passed to the transformer encoder where each patch is weighted differently according to its significance. The encoder is made up of Multi-Head Self-Attention layers (MSP), Multi-Layer Perceptron layer (MLP), and Layer Norms (LN). For this study, we have used ViT-B32 version provided by the ViT_Keras library.

*F. Xception*

Xception is another CNN-based architecture that relies solely on depth-wise separable convolutions. It was introduced by F. Chollet [16] who also happens to be the creator of Keras library. It is considered an improvement over the InceptionV3 network. It outperforms InceptionV3 by utilizing fewer parameters. The fact that it runs on depth-wise separable convolutions makes it ideologically similar to MobileNets. However, the motivation for developing these two architectures was very distinct. Xception was built to keep the model's accuracy in focus, while MobileNets emphasize more on light weightiness property of the model. We have used the pre-trained version of the model available with the Keras application library.

## V. EXPERIMENTAL SETUP

This section discusses the experimental setup and the configurations used while training and evaluating the models. A glimpse of the methodology employed is given in "Fig. 4". The experimentation was performed on a competent GPU i.e., Nvidia RTX A5000 24GB. The images retrieved from the dataset were of different dimensions that have to be converted into a dimension of (224,224,3) as all models selected for the comparative study were pre-trained on similar sizes of images. This process was covered under the data pre-processing step. Data augmentation is the next step in line. Here, we tried to induce randomness into the data set by transposing, flipping, rotating, and saturating the pixels of images. Augmenting data in general cases is known to improve learning performance and reduce problems like overfitting [17]. The pre-processed and augmented data is then fed to the selected models, discussed in section IV, individually for training purpose. The evaluation of all the models were done using 5-fold cross-validation [18]. It was done to get a better statistical idea of the performance of individual model. Model's mean accuracy and mean prediction time were the primary parameters under observation. Where accuracy was calculated using the predefined function available in Keras.metrics, the prediction was calculated by noting TensorFlow's model.predict for the entire test set and averaging it to get the time required for a single prediction.

Now, we discuss the configurational aspect of the training process involved. The same configurations were used with all the models discussed to facilitate a non-partial performance comparison. Popular python frameworks Keras and TensorFlow 2.0 were used to perform the study. The process initiates with loading of the pixel values of the images along with their respective classes using ImageDataGenerator utility of Keras into 32-size batches.

Various standardization and normalization techniques were applied to these batches. These include (1) rescaling the pixel intensity from 0-255 to 0-1, (2) performing sample wise centre, and (3) sample wise standard normalization.

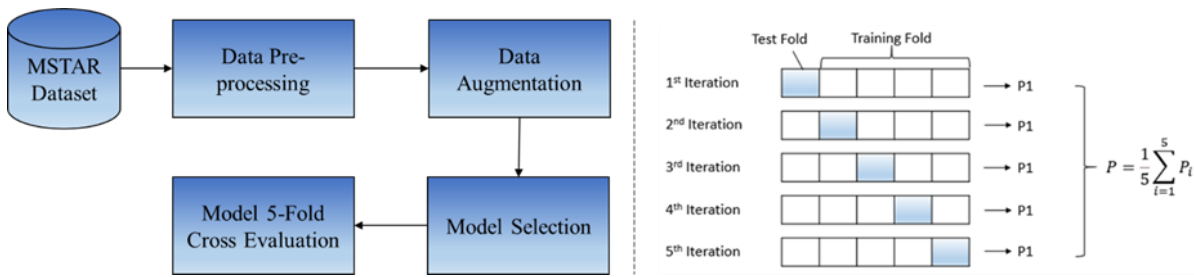

Fig. 4. ConvNext block

Here, the second step shifts the origin of intensity distribution by subtracting with its sample's mean $x_i = x_i - \sum \frac{x_i}{n}$. The third step scales the intensity by dividing the value by its sample's standard deviation $x_i = \frac{x_i}{\sqrt{\frac{\sum |x-\bar{x}|^2}{n}}}$. It helps in reducing the chances of exploding gradient problems along with ensuring a faster convergence of the model [19]. The batch pixel intensity is visualized in the "Fig. 5" (before and after normalization).

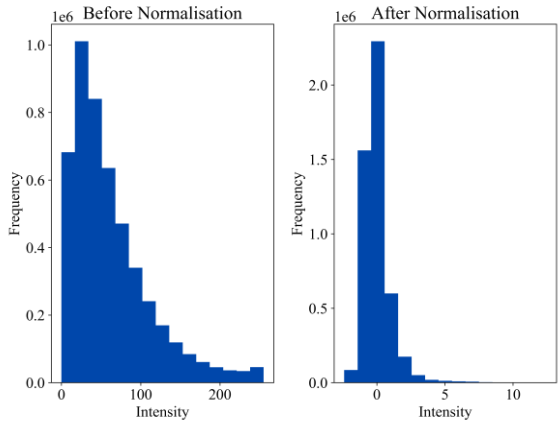

Fig. 5. Histogram plot of batch pixel intensities.

### A. Model compilation

The loss function used here was cross-entropy loss, since the problem was of a categorical type. Label smoothing option was set to a factor of 0.2 to make learning more generalized. Optimizer is also one of the other important compilation parameters. Here we have used rectified adam (RAdam) for the case. RAdam is proved to be better in handling large variance problems of Adam optimizer for the limited dataset at an early stage of training [20]. The learning rate (lr) required by the optimizer was provided in a discriminative way. A smaller lr (1*e-5) was chosen to fine-tune the pre-trained models on the target dataset, while a larger lr (1*e-3) was used for the last prediction dense layer. The rationale behind this was to preserve more information from the feature vector of the trained models by slowly updating it to adapt to the target problem, and also to let the randomly initialized dense layer learn at a faster pace to decrease overall model training time. The lr rate was also dynamically reduced using modified ReduceLROnPlateau callback utility from Keras. It reduces the lr by the factor of 0.33 for feature vector and 0.1 for top dense layer if the validation accuracy does not improve for 2 consecutive epochs. The other callbacks used in the training were EarlyStopping (built-in) to stop model from overtraining, GC callback to handle garbage collection and memory leaks and a Time callback to save intermediate training variables like epoch count and epoch time for every training fold.

## VI. RESULTS

The results obtained from the study are detailed in this section. This study is an attempt to explore the suitability of current state-of-the-art models introduced in the domain of computer vision for SAR target classification. Here, the results of DenseNet121 are used as the baseline for the comparative study. First, we take a glance at the training process of these models. "Fig. 6" & "Fig. 7" represent the validation loss and accuracy curves of all the models. These curves broadly indicate that all models have performed well compared with the baseline. The baseline's validation curves were quite bumpy indicating a less stable learning process. The other models, on the other hand, have performed quite smoothly. On a finer look, we may observe that ConvNext model has the smoothest curve and has converged in the least number of epochs. It is followed by BiT and ViT.

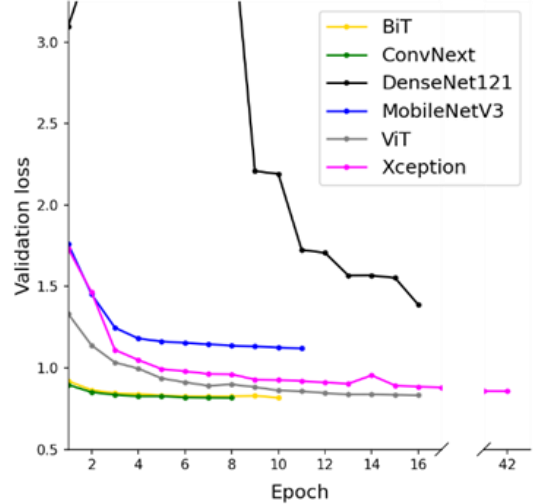

Fig. 6. Loss curves of trained models.

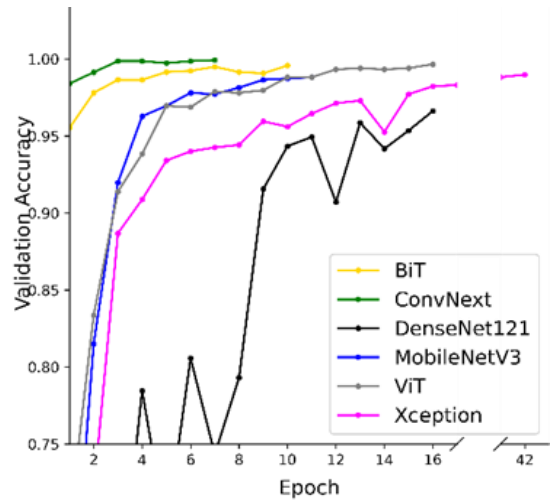

Fig. 7. Loss curves of trained models.

The "Table I" summarises the performance of individual models after 5-fold cross-validation process. Model accuracies and losses at various steps (training, validation, and test) were captured. These values represent the mean of all the 5 folds that the models have gone through. It may be noted from these values that ConvNext with test accuracy (mean) 99.87± 0.09% was the model that outperformed the rest of the models. It was found to have about 4% more test accuracy than the baseline model. Again, it was seen to be closely trailed by BiT and ViT respectively.

TABLE I. THE LOSS AND ACCURACY OF THE TRAINED MODELS.

| Target classification models | Training loss mean (Std) | Training accuracy mean (Std) | Validation loss mean (Std) | Validation accuracy mean (Std) | Test loss mean (Std) | Test accuracy mean (Std) |
|---|---|---|---|---|---|---|
| BiT | 0.8106 (0.0023) | 0.9994 (0.0005) | 0.8182 (0.0039) | 0.9971 (0.0013) | 0.8170 (0.0031) | 0.9971 (0.0015) |
| ConvNext | 0.8174 (0.0110) | 0.9993 (0.00105) | 0.8195 (0.0105) | 0.9995 (0.0005) | 0.8193 (0.0108) | **0.9987** (0.0009) |
| DenseNet121 | 1.5688 (0.2681) | 0.9711 (0.0040) | 1.6320 (0.2807) | 0.9575 (0.0052) | 1.6680 (0.4253) | 0.9583 (0.0096) |
| MobileNetV3 | 1.0979 (0.0086) | 0.9957 (0.0018) | 1.1100 (0.0070) | 0.9906 (0.0016) | 1.1116 (0.0045) | 0.9899 (0.0016) |
| ViT | 0.8329 (0.0108) | 0.9949 (0.0023) | 0.8340 (0.0099) | 0.9951 (0.0026) | 0.8352 (0.0076) | 0.9941 (0.0020) |
| Xception | 0.8497 (0.0104) | 0.9951 (0.0023) | 0.8767 (0.0116) | 0.9850 (0.0031) | 0.8800 (0.0140) | 0.9792 (0.0070) |

The study also tried to include the second most important dimension i.e., prediction time of comparison. The "Fig. 8" represents a bubble chart where the y-axis denotes the mean test accuracies of the model (in %), and the x-axis denotes the mean prediction time required to classify a single data point (in milliseconds). The size of the bubble represents a separate dimension i.e., the training time required by individual models. It is calculated by the average time required to complete a training epoch (Et) multiplied by the average number of epochs taken by a model to converge (En). Thus the training time would be "Et*En". It is used as a comparative quantity, and we are not interested in its absolute numbers here.

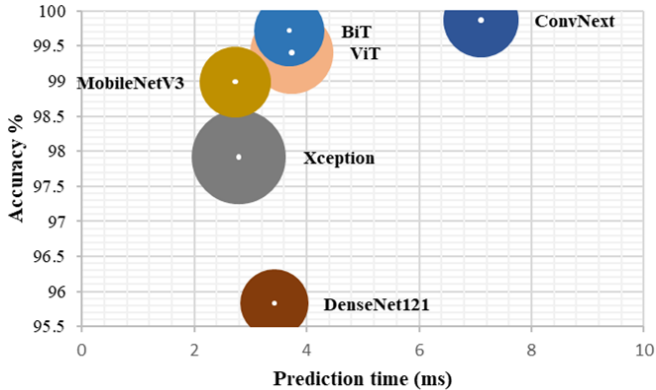

Fig. 8. Accuracy vs. prediction time plot, where bubble size represents the training time of a model.

Now, we will try to comprehend the "Fig. 8". The observation from "Table I" inferred that ConvNext was the best model so far in terms of test accuracy. But, the "Fig. 8" shows the other side of the picture. The accuracy achieved by ConvNext comes with the cost of increased prediction time. It may be seen to have almost double the prediction time when compared to the fastest in the figure i.e., MobileNetV3. Even the baseline model has performed better than ConvNext in terms of prediction time. One probable explanation for the observation could be that the complicated architecture of the ConvNext allows it to learn better features from the target, which on the other hand complicates the prediction process and increases its prediction time as well. If responsiveness is considered the primary objective of a mission, then one may recommend (from the "Fig. 8") the MobileNetV3 model for the task. However, if the objective requires a blend of high accuracy and good response time, then BiT would be an obvious choice. The other parameter that can be noted from the same figure is bubble size (training time). DenseNet121 (baseline) was seen to have the least training time while Xception had the largest training time among all.

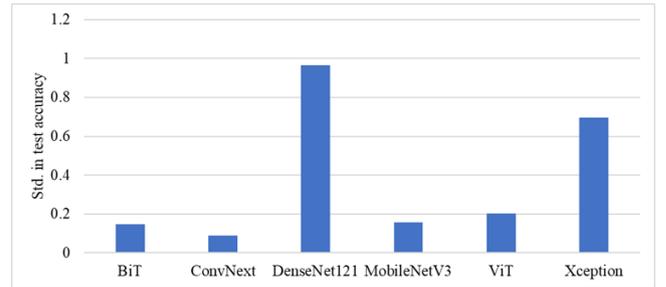

Fig. 9. Standard deviation from mean test accuracy.

The "Fig. 9" is plotted to show the variation in test accuracy percentage as models are tested on every data point in successive validation folds. Standard deviation is the statistical measure used to track this variation. ConvNext was observed to have the least deviation from mean test accuracy followed by BiT. This may be interpreted as it has shown the least input sequence dependency in the training phase and was found to be the most input resilient when compared with other models. However, the difference here is not much prominent. Two probable reasons could be that the dataset used contained a limited number of images and that the images were too simplistic/ idealistic (contained only one object per image). The results may turn very useful when considering a bigger and more complex dataset of the task.

VII. CONCLUSION

The paper explores the application of deep learning models in the domain of SAR image classification. It produces an in-depth comparative analysis of the different models applied to SAR image classification. Apart from accuracy, the study tried to include other important dimension of prediction time. To add further, the dependency between a model's training and its input sequence was also analyzed. The results revealed that ConvNext was the most accurate classification model. It was also found to be most input resilient. However, it lacked drastically on the parameter of responsiveness. BiT appeared to be the best choice when considering a blend of accuracy and responsiveness. It has also shown good input resilient characteristics. The results thus present a useful insight on application of deep learning models for SAR image classification. The finding of the study could be utilized by agencies and departments working in strategic sectors to develop a state-of-the-art SAR system. All models compared are in their vanilla state and hence have the scope of further improvement in their performance by considering other domain-specific customizations.

REFERENCES

[1] J. J. Thiagarajan, K. N. Ramamurthy, P. Knee, A. Spanias, and V. Berisha, "Sparse representations for automatic target classification in SAR images," in *2010 4th international symposium on communications, control and signal processing (ISCCSP)*, 2010, pp. 1–4.

[2] R. K. Mehra, R. B. Ravichandran, and A. Srivastava, "MSTAR target classification using Bayesian pattern theory," in *Algorithms for Synthetic Aperture Radar Imagery V*, 1998, vol. 3370, pp. 675–684.

[3] M. L. Bryant and F. D. Garber, "SVM classifier applied to the MSTAR public data set," in *Algorithms for Synthetic Aperture Radar Imagery VI*, 1999, vol. 3721, pp. 355–360.

[4] Q. Zhao and J. C. Principe, "Support vector machines for SAR automatic target recognition," *IEEE Transactions on Aerospace and Electronic Systems*, vol. 37, no. 2, pp. 643–654, 2001.

[5] Y. Yang, Y. Qiu, and C. Lu, "Automatic target classification-experiments on the MSTAR SAR images," in *Sixth international conference on software engineering, artificial intelligence, networking and parallel/distributed computing and first ACIS international workshop on self-assembling wireless network*, 2005, pp. 2–7.

[6] A. Profeta, A. Rodriguez, and H. S. Clouse, "Convolutional neural networks for synthetic aperture radar classification," in *Algorithms for synthetic aperture radar imagery XXIII*, 2016, vol. 9843, pp. 185–194.

[7] C. Coman and others, "A deep learning sar target classification experiment on mstar dataset," in *2018 19th international radar symposium (IRS)*, 2018, pp. 1–6.

[8] G. M. Weiss and F. Provost, "The effect of class distribution on classifier learning: an empirical study," Rutgers University, techreport, 2001.

[9] Kolesnikov *et al.*, "Big transfer (bit): General visual representation learning," in *European conference on computer vision*, 2020, pp. 491–507.

[10] Z. Liu, H. Mao, C.-Y. Wu, C. Feichtenhofer, T. Darrell, and S. Xie, "A convnet for the 2020s," in *Proceedings of the IEEE/CVF Conference on Computer Vision and Pattern Recognition*, 2022, pp. 11976–11986.

[11] G. Huang, Z. Liu, L. Van Der Maaten, and K. Q. Weinberger, "Densely connected convolutional networks," in *Proceedings of the IEEE conference on computer vision and pattern recognition*, 2017, pp. 4700–4708.

[12] G. Howard *et al.*, "Mobilenets: Efficient convolutional neural networks for mobile vision applications," *arXiv preprint arXiv:1704.04861*, 2017.

[13] Howard *et al.*, "Searching for mobilenetv3," in *Proceedings of the IEEE/CVF international conference on computer vision*, 2019, pp. 1314–1324.

[14] Vaswani *et al.*, "Attention is all you need," *Advances in neural information processing systems*, vol. 30, 2017.

[15] Dosovitskiy *et al.*, "An image is worth 16x16 words: Transformers for image recognition at scale," *arXiv preprint arXiv:2010.11929*, 2020.

[16] F. Chollet, "Xception: Deep learning with depthwise separable convolutions," in *Proceedings of the IEEE conference on computer vision and pattern recognition*, 2017, pp. 1251–1258.

[17] L. Perez and J. Wang, "The effectiveness of data augmentation in image classification using deep learning," *arXiv preprint arXiv:1712.04621*, 2017.

[18] D. Anguita, L. Ghelardoni, A. Ghio, L. Oneto, and S. Ridella, "The 'K'in K-fold cross validation," in *20th European Symposium on Artificial Neural Networks, Computational Intelligence and Machine Learning (ESANN)*, 2012, pp. 441–446.

[19] C. M. Bishop and others, *Neural networks for pattern recognition*. Oxford university press, 1995.

[20] L. Liu *et al.*, "On the variance of the adaptive learning rate and beyond," *arXiv preprint arXiv:1908.03265*, 2019.